\documentclass{ws-procs975x65}

\begin{document}

\title{MG13 proceedings: 
  Construction of gauge-invariant variables for
  linear-order metric perturbations on an arbitrary background
  spacetime}

\author{
  Kouji Nakamura
}

\address{TAMA Project, Optical and Infrared Astronomy Division,\\
  National Astronomical Observatory of Japan\\
  2-21-1 Osawa, Mataka, Tokyo, 181-8588, Japan\\
  E-mail: kouji.nakamura@nao.ac.jp}

\begin{abstract}
  An outline of a proof of the decomposition of the linear
  metric perturbation into gauge-invariant and gauge-variant
  parts on an arbitrary background spacetime is discussed
  through an exlicit construction of gauge-invariant and
  gauge-variant parts.
  Although this outline is incomplete, yet, due to our
  assumptions, we propose a conjecture which states that the
  linear metric perturbation is always decomposed into its
  gauge-invariant and gageu-variant parts.
  If this conjecture is true, we can develop the higher-order
  gauge-invariant perturbation theory on an arbitrary background
  spacetime.
\end{abstract}


\bodymatter

\section{Introduction}
\label{sec:intro}


As well-known, general relativity is based on general covariance
and the ``gauge degree of freedom'', which is unphysical degree
of freedom of perturbations, arises due to this general
covariance.
Furthermore, gauge-transformation rules for higher-order
perturbations are very complicated.
So, it is worthwhile to investigate higher-order gauge-invariant
perturbation theory from a general point of view.


According to this motivations, we have been formulating the
higher-order general-relativistic gauge-invariant perturbation
theory~\cite{kouchan-gauge-inv-O28-09}.
These works are based on the single assumption that {\it we
  already know the procedure to find gauge-invariant variables
  for linear-order metric perturbations}.
(Conjecture \ref{conjecture:decomposition-conjecture-O28-09} in
this article) and our formulation is well-defined except for
this assumption.


The main purpose of this article is to give a brief outline of a 
proof of this assumption~\cite{K.Nakamura-2011-full-paper-O28-09}.


\section{Perturbations in general relativity and gauge-invariant
variables}
\label{sec:Perturbations_in_general_relativity}


Here, we concentrate on the second-kind gauge in
perturbation theories with general
covariance~\cite{kouchan-gauge-inv-O28-09}.
In perturbation theories, we always treat two spacetime
manifolds.
One is the physical spacetime ${\cal M}_{\lambda}$ which is our
nature itself and another is the background spacetime 
${\cal M}_{0}$ which is prepared by hand for perturbative
analyses.
{\it The gauge choice of the second kind} is the point
identification map ${\cal X}_{\lambda}$ $:$ 
${\cal M}_{0}\mapsto{\cal M}_{\lambda}$.
{\it The gauge transformation of the second kind} is a change
${\cal X}_{\lambda}\rightarrow{\cal Y}_{\lambda}$ of this
identification.


Once we specify a gauge choice ${\cal X}_{\lambda}$, we can
define perturbations of a physical variable $\bar{Q}_{\lambda}$
using the pulled-back ${\cal X}_{\lambda}^{*}\bar{Q}$ of
$\bar{Q}_{\lambda}$.
${\cal X}_{\lambda}^{*}\bar{Q}$ is expanded as  
\begin{equation}
  \label{eq:Bruni-35-O28-09}
  {}^{\cal X}\!Q
  :=
  \left.{\cal X}^{*}_{\lambda}\bar{Q}_{\lambda}\right|_{{\cal M}_{0}} 
  =
  Q_{0}
  + \lambda {}^{(1)}_{\;\cal X}\!Q
  + \frac{1}{2} \lambda^{2} {}^{(2)}_{\;\cal X}\!Q
  + O(\lambda^{3}).
\end{equation}
Here, ${}^{(1)}_{\;\cal X}\!Q$ (${}^{(2)}_{\;\cal X}\!Q$) are
the first-order (second-order) perturbation of $\bar{Q}_{\lambda}$.


The diffeomorphism 
$\Phi_{\lambda}:=({\cal X}_{\lambda})^{-1}\circ{\cal Y}_{\lambda}$ 
is the map $\Phi_{\lambda}$ $:$ ${\cal M}_{0}$ $\rightarrow$
${\cal M}_{0}$ and does change the point identification. 
So, $\Phi_{\lambda}$ is the {\it gauge transformation}
$\Phi_{\lambda}$ $:$ ${\cal X}_{\lambda}$ $\rightarrow$ 
${\cal Y}_{\lambda}$ and the induced pull-back operates as 
${}^{\cal Y}Q_{\lambda}=\Phi^{*}_{\lambda} {}^{\cal X}Q_{\lambda}$. 
The generic Taylor expansion~\cite{kouchan-gauge-inv-O28-09}
leads the order-by-order gauge-transformation rules for the
perturbations as
\begin{eqnarray}
  \label{eq:Bruni-47-one-O28-09}
  {}^{(1)}_{\;{\cal Y}}\!Q - {}^{(1)}_{\;{\cal X}}\!Q = 
  {\pounds}_{\xi_{(1)}}Q_{0}, \quad
  {}^{(2)}_{\;\cal Y}\!Q - {}^{(2)}_{\;\cal X}\!Q = 
  2 {\pounds}_{\xi_{(1)}} {}^{(1)}_{\;\cal X}\!Q 
  +\left\{{\pounds}_{\xi_{(2)}}+{\pounds}_{\xi_{(1)}}^{2}\right\} Q_{0}.
\end{eqnarray}
where $\xi_{(1)}^{a}$ and $\xi_{(2)}^{a}$ are the generators of
$\Phi_{\lambda}$.


We call the $k$th-order perturbation ${}^{(k)}_{{\cal X}}\!Q$ is
{\it gauge invariant} iff ${}^{(k)}_{\;\cal X}\!Q = {}^{(k)}_{\;\cal Y}\!Q$
for any gauge choice ${\cal X}_{\lambda}$ and
${\cal Y}_{\lambda}$~\cite{kouchan-gauge-inv-O28-09}.


Through these setup, we first consider the metric perturbation
to construct gauge-invariant variables for higher-order
perturbations~\cite{kouchan-gauge-inv-O28-09}.
The pulled-back metric ${\cal X}^{*}_{\lambda}\bar{g}_{ab}$ is
expanded as Eq.~(\ref{eq:Bruni-35-O28-09}): 
${\cal X}^{*}_{\lambda}\bar{g}_{ab}$ $=$ $g_{ab}$ $+$  
$\lambda {}_{{\cal X}}\!h_{ab}$ $+$ 
$(\lambda^{2}/2){}_{{\cal X}}\!l_{ab}$ $+$ $O^{3}(\lambda)$, 
where $g_{ab}$ is the metric on ${\cal M}_{0}$.
Our starting point of the construction of gauge-invariant
variables is the following assumption for $h_{ab}$:
\begin{conjecture}
  \label{conjecture:decomposition-conjecture-O28-09}
  If there is a symmetric tensor field $h_{ab}$ of the second
  rank, whose gauge transformation rule is 
  ${}_{{\cal Y}}\!h_{ab}$ $-$ ${}_{{\cal X}}\!h_{ab}$ $=$
  ${\pounds}_{\xi_{(1)}}g_{ab}$, then there exist a tensor field
  ${\cal H}_{ab}$ and a vector field $X^{a}$ such that $h_{ab}$
  is decomposed as $h_{ab}$ $=:$ ${\cal H}_{ab}$ $+$
  ${\pounds}_{X}g_{ab}$, where ${\cal H}_{ab}$ and $X^{a}$ are
  transformed as ${}_{{\cal Y}}\!{\cal H}_{ab}$ $-$ 
  ${}_{{\cal X}}\!{\cal H}_{ab}$ $=$ $0$, 
  ${}_{\quad{\cal Y}}\!X^{a}$ $-$ ${}_{{\cal X}}\!X^{a}$ $=$
  $\xi^{a}_{(1)}$ under the gauge transformation
  (\ref{eq:Bruni-47-one-O28-09}), respectively.
\end{conjecture}
In this conjecture, ${\cal H}_{ab}$ and $X^{a}$ are 
{\it gauge-invariant} and {\it gauge-variant} parts of the
perturbation $h_{ab}$, respectively.
If we accept Conjecture
\ref{conjecture:decomposition-conjecture-O28-09}, we can 
recursively define gauge-invariant variables for higher-order
perturbations~\cite{kouchan-gauge-inv-O28-09}.


\section{An outline of a proof of Conjecture
  \ref{conjecture:decomposition-conjecture-O28-09}} 
\label{sec:An_outline_of_a_proof_of_Conjecture}


To prove Conjecture
\ref{conjecture:decomposition-conjecture-O28-09}, we assume that
the background spacetimes ${\cal M}_{0}$ admit ADM
decomposition, whose metric is given by $g_{ab}$ $=$
$-\alpha^{2}(dt)_{a}(dt)_{b}$ $+$
$q_{ij}(dx^{i}+\beta^{i}dt)_{a}(dx^{j}+\beta^{j}dt)_{b}$. 
We decompose of the components $\{h_{ti},h_{ij}\}$ of $h_{ab}$
as 
\begin{eqnarray}
  &&
  h_{ti}
  =:
  D_{i}h_{(VL)} + h_{(V)i}
  - \frac{2}{\alpha} \left(
    D_{i}\alpha 
    - \beta^{k}K_{ik}
  \right) \left(
    h_{(VL)}
    - \Delta^{-1}D^{k}\partial_{t}h_{(TV)k}
  \right)
  \nonumber\\
  && \quad\quad\quad
  - \frac{2}{\alpha} M_{i}^{\;\;k} h_{(TV)k}
  \label{eq:hti-decomp-alternative-O28-09}
  , \\
  &&
  h_{ij}
  =:
  \frac{1}{n} q_{ij} h_{(L)} +
  D_{i}h_{(TV)j}+D_{j}h_{(TV)i}-\frac{2}{n}q_{ij}D^{k}h_{(TV)k}
  + h_{(TT)ij} 
  \nonumber\\
  && \quad\quad\quad
  + \frac{2}{\alpha} K_{ij} \left(
    h_{(VL)}
    - \Delta^{-1}D^{k}\partial_{t}h_{(TV)k}
  \right)
  - \frac{2}{\alpha} K_{ij} \beta^{k} h_{(TV)k}
  \label{eq:hij-decomp-alternative-O28-09}
  , \\
  && D^{i}h_{(V)i} = 0, \quad q^{ij} h_{(TT)ij} = 0 = D^{i}h_{(TT)ij}
  \label{eq:hti-hij-decomp-conditions-O28-09}
  .
\end{eqnarray}
where $M_{i}^{\;\;j}$ is defined by $M_{i}^{\;\;j}$ $:=$ $-$
$\alpha^{2}K^{j}_{\;\;i}$ $+$ $\beta^{j}\beta^{k} K_{ki}$ $-$
$\beta^{j}D_{i}\alpha$ $+$ $\alpha D_{i}\beta^{j}$. 
Here, $K_{ij}$ is the extrinsic curvature and $D_{i}$ is the
covariant derivative associate with the metric $q_{ij}$ on
$t=const$ hypersurfaces.


Here, we assumeed the existence of Green functions of the elliptic
derivative operators $\Delta:=D^{i}D_{i}$ and ${\cal F}$ $:=$
$\Delta$ $-$
$\frac{2}{\alpha}\left(D_{i}\alpha-\beta^{j}K_{ij}\right)D^{i}$ 
$-$
$2D^{i}\left\{\frac{1}{\alpha}\left(D_{i}\alpha-\beta^{j}K_{ij}\right)\right\}$, 
and the existence and the uniqueness of the solution $A_{i}$ to
the equation 
\begin{eqnarray}
  &&
  {\cal D}_{j}^{\;\;k}A_{k}
  + D^{m}\left[
    \frac{2}{\alpha} \tilde{K}_{mj} \left\{ \frac{}{}
      {\cal F}^{-1}D^{k}\left(
        \frac{2}{\alpha} M_{k}^{\;\;l} A_{l}
        - \partial_{t}A_{k}
      \right)
      - \beta^{k} A_{k}
      \frac{}{}
    \right\}
  \right]
  =
  L_{j}
  \label{eq:K.Nakamura-2010-note-IV-92-O28-09}
\end{eqnarray}
for given a vector field $L_{j}$.
We note that the relations
(\ref{eq:hti-decomp-alternative-O28-09})--(\ref{eq:hti-hij-decomp-conditions-O28-09})
are invertible if we accept these three assumptions. 
These assumptions also imply that we have ignored perturbative
modes which belong to the kernel of the above derivative 
operators and trivial solutions to
Eq.~(\ref{eq:K.Nakamura-2010-note-IV-92-O28-09}).
We call these modes as {\it zero modes}.
The issue on the treatments of these zero modes is
called {\it zero-mode problem}, which is a remaining problem in
our formulation.


Due to
Eqs.~(\ref{eq:hti-decomp-alternative-O28-09})--(\ref{eq:hti-hij-decomp-conditions-O28-09}),
the gauge-transformation rule ${}_{{\cal Y}}\!h_{ab}$ $-$
${}_{{\cal X}}\!h_{ab}$ $=$ ${\pounds}_{\xi_{(1)}}g_{ab}$ leads
\begin{eqnarray}
  &&
  {}_{{\cal Y}}h_{(VL)} - {}_{{\cal X}}h_{(VL)}
  =
  \xi_{t}
  + \Delta^{-1}D^{k}\partial_{t}\xi_{k}
  \label{eq:K.Nakamura-2010-note-B-44-O28-09}
  , \;
  {}_{{\cal Y}}h_{(V)i} - {}_{{\cal X}}h_{(V)i}
  =
    \partial_{t}\xi_{i}
  - D_{i}\Delta^{-1}D^{k}\partial_{t}\xi_{k}
  , \nonumber\\
  &&
  {}_{{\cal Y}}h_{(L)} - {}_{{\cal X}}h_{(L)}
  =
  2 D^{i}\xi_{i}
  \label{eq:K.Nakamura-2010-note-B-46-O28-09}
  , \;
  {}_{{\cal Y}}h_{(TV)l} - {}_{{\cal X}}h_{(TV)l}
  = \xi_{l}
  ,\;
  {}_{{\cal Y}}h_{(TT)ij} - {}_{{\cal X}}h_{(TT)ij} = 0.
  \nonumber
\end{eqnarray}
These yield the gauge-variant part $X_{a}$ in Conjecture
\ref{conjecture:decomposition-conjecture-O28-09} is given by
$X_{a}$ $=$ $X_{t}(dt)_{a}$ $+$ $X_{i}(dx^{i})_{a}$ with $X_{i}$
$:=$ $h_{(TV)i}$ and $X_{t}$ $:=$ $h_{(VL)}$ $-$
$\Delta^{-1}D^{k}\partial_{t}h_{(TV)k}$.
Using the variables $X_{t}$ and $X_{i}$, we can construct
gauge-invariant variables for $h_{ab}$ as
\begin{eqnarray}
  \label{eq:K.Nakamura-2010-note-B-62-O28-09}
  - 2 \Phi
  &:=&
  h_{tt}
  + \frac{2}{\alpha}\left(
    \partial_{t}\alpha + \beta^{i}D_{i}\alpha - \beta^{j}\beta^{i}K_{ij}
  \right) X_{t}
  - 2 \partial_{t}X_{t}
  \nonumber\\
  &&
  + \frac{2}{\alpha} \left(
    \beta^{i}\beta^{k}\beta^{j}K_{kj} - \beta^{i}\partial_{t}\alpha 
    + \alpha q^{ij}\partial_{t}\beta_{j}
    + \alpha^{2}D^{i}\alpha
    - \alpha\beta^{k}D^{i}\beta_{k}
    - \beta^{i}\beta^{j}D_{j}\alpha
  \right)
  X_{i}
  , \nonumber\\
  \label{eq:K.Nakamura-2010-note-B-64-O28-09}
  - 2 n \Psi
  &:=&
  h_{(L)} - 2 D^{i}X_{i}
  , \quad
  \nu_{i}
  :=
  h_{(V)i}
  - \partial_{t}X_{i}
  + D_{i}\Delta^{-1}D^{k}\partial_{t}X_{k}
  , \quad
  \chi_{ij} := h_{(TT)ij},
  \nonumber
\end{eqnarray}
where $n$ is the dimension of the $t=const$ hypersurface.
The representations of the original components of $h_{ab}$
in terms of these gauge-invariant variables, $X_{t}$, and
$X_{i}$ yield the assertion of Conjecture 
\ref{conjecture:decomposition-conjecture-O28-09}. $\Box$


\section{Discussion}
\label{sec:discussion}


Due to the above proof of Conjecture
\ref{conjecture:decomposition-conjecture-O28-09}, we almost 
completed our formulation of general-relativistic higher-order
gauge-invariant perturbation theories.
This indicates the possibility of the wide applications of our
formulation.
Although our arguments do not include zero modes and these also
have their physical
meaning~\cite{K.Nakamura-2011-GREssey-O28-09}, 
we propose Conjecture
\ref{conjecture:decomposition-conjecture-O28-09} as an
conjecture.



\end{document}